\documentclass{webofc}
\usepackage[varg]{txfonts}   

\usepackage{graphicx}
\usepackage{amsmath}
\usepackage{graphicx}
\usepackage{amsmath}
\newcommand{\beq}{\begin{equation}}
\newcommand{\eeq}{\end{equation}}
\newcommand{\bqa}{\begin{eqnarray}}
\newcommand{\eqa}{\end{eqnarray}}

\def\square{\vcenter{\vbox{\hrule height.4pt
          \hbox{\vrule width.4pt height4pt
          \kern4pt\vrule width.3pt}\hrule height.4pt}}}

\begin{document}
\title{Phases and condensates in zero-temperature 
QCD at finite $\mu_I$ and $\mu_S$}
%
%

\author{\firstname{Jens O} \lastname{Andersen}\fnsep\thanks{\email{andersen@tf.phys.ntnu.no}} 
}

\institute{
Department of Physics,  Faculty of Natural Sciences, 
Norwegian University of Science and Technology, Høgskoleringen 5, Trondheim, N-7491, Norway, \\ 
Niels Bohr International Academy, Blegdamsvej 17,
DK-2100 Copenhagen, Denmark}

\abstract{%
I discuss pion and koan condensation and the 
the properties of the phases of QCD at finite isospin
chemical potential $\mu_I$
and strangeness chemical potential $\mu_S$ at zero temperature
using three-flavor chiral perturbation theory.
Electromagnetic effects are included in the calculation of the phase diagram, which implies that the 
charged meson condensed phases become superconducting phases of QCD
with a massive photon via the Higgs mechanism.
Without electromagnetic effects, we show results for the light quark condensate and the pion condensate as functions of $\mu_I$ at next-to-leading (NLO) order
in the low-energy expansion. The results are compared with recent lattice
simulations and by including the NLO corrections, one obtains very
good agreement.

}
\maketitle
\section{Introduction}
\label{intro}
\begin{figure}[htb!]
\centering
        \includegraphics[scale=0.6]{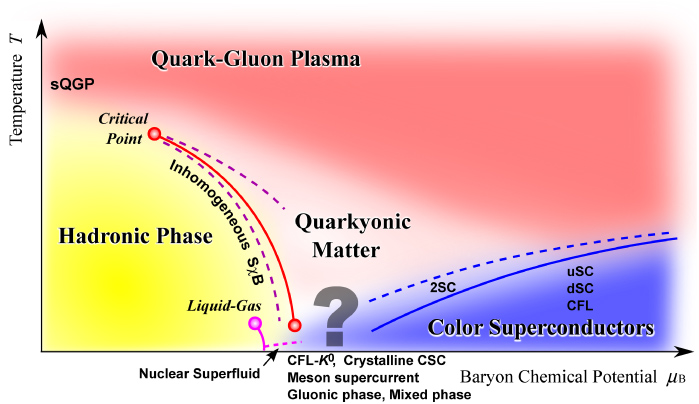}
\caption{Phase diagram of QCD in the $\mu_B$--$T$ plane. Figure from Ref.~\cite{hatsuda}.}
\label{fase0}
\end{figure}

In this talk, I would like to discuss various aspects of the
phases of QCD at zero temperature, but finite isospin and strangeness density.
However, before I do that, I would like to briefly comment on the
QCD phase diagram as it is normally presented, namely in the $\mu_B$--$T$ plane.
It is shown in Fig.~\ref{fase0}, borrowed from Ref.~\cite{hatsuda}. 
Few of the results for the phase diagram are
rigorous in the sense that they are obtained from first principles, rather they
are obtained by model calculations. For asymptotically high temperatures and zero
baryon chemical potential, we know that QCD is in a quark-gluon plasma phase
consisting of weakly interacting deconfined quarks and gluons.
Similarly, we know that at asymptotically high baryon density and zero temperature,
QCD is in the color-flavor locked phase arising from an attracting channel of
one-gluon exchange and the resulting instability of the Fermi surface.
From lattice simulations, we know that there is a cross-over transition for
$\mu_B=0$ at a temperature of around 155 MeV.
For low temperatures and large chemical potentials, the infamous sign problem,
prohibits the use of standard Monte Carlo techniques to study the properties
of QCD. One must therefore resort to low-energy models
such as the NJL model and the quark-meson model. Over the past two decades,
a huge amount of work has been done to map out the phase diagram.

The situation is even more complex than this since, instead of using a common quark chemical 
potential for all quarks, one can introduce a separate chemical potential $\mu_f$ for
each flavor. For three flavors, we use either $\mu_u$, $\mu_d$, and $\mu_s$
or the baryon chemical potential $\mu_B$, the  isospin chemical potential $\mu_I$,
and strangeness chemical potential $\mu_S$ defined as
\bqa
\mu_B={3\over2}(\mu_u+\mu_d)\;,
\hspace{1cm}
\mu_I={1\over2}(\mu_u-\mu_d)\;,
\hspace{1cm}
\mu_S={1\over2}(\mu_u+\mu_d-2\mu_s)\;.
\eqa
For $\mu_B=\mu_S=0$ but nonzero $\mu_I$, 
one can carry out Monte Carlo simulations using standard techniques
since the fermion determinant is real in this case and consequently there is no
sign problem. This opens up the possibility to study charged pion condensation 
on the lattice and confront it with results from low-energy effective theories.
In this talk, I will discuss pion condensation for two and three
flavors using chiral perturbation theory ($\chi$PT) as a low-energy effective theory
for QCD and show results for the light quark and pion condensates as a function of $\mu_I$ with $\mu_B=\mu_S=0$. The results will be compared
to recent high-precision lattice simulations~\cite{gergo,gergo1,gergo2,gergo3}. 
I will also discuss the phase
diagram and meson condensation for three flavors
in the $\mu_I$--$\mu_S$ plane at zero temperature.

\begin{figure}[htb!]
\centering
        \includegraphics[scale=0.3]{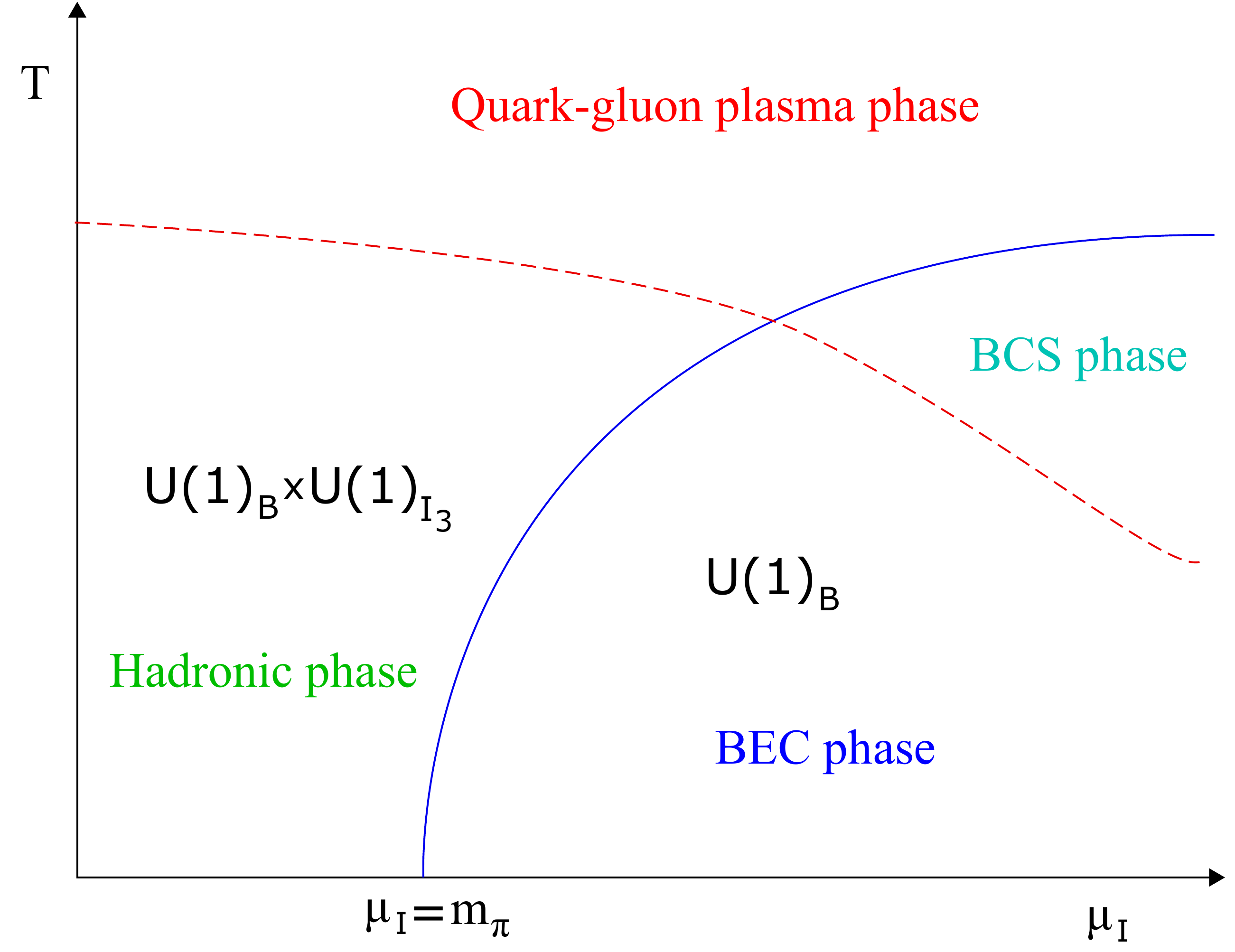}
\caption{Phase diagram of QCD in the $\mu_I$--$T$ plane.}
\label{fase}
\end{figure}

In Fig.~\ref{fase}, we sketch the phase diagram of two-flavor QCD in the 
the $T$--$\mu_I$ plane. In the lower left region, we have the hadronic phase
where chiral symmetry is broken and quarks are confined. 
As the temperature increases, one enters the
quark-gluon plasma phase. Along the $\mu_I$ axis, there is a transition from the
hadronic phase to a Bose-condensed phase of charged pions. In this phase, the
$U(1)_{I_3}$ symmetry is broken giving rise to a massless Goldstone boson, which
is a mixture of $\pi^+$ and $\pi^-$. For large isospin chemical and low temperature,
one expects that quarks are the relevant degrees of freedom rather than pions~\cite{son} .
The Fermi surface that exists when the interactions are turned off, is rendered
unstable once they are turned on, since they are attractive. The system is then described
in terms of loosely bound Cooper pairs instead of tightly bound pions. Since the
symmetry breaking pattern is the same, there is a cross-over transition rather
than a true phase transition between the BEC and the BCS phases~\cite{son}.
 
\section{$\chi$PT at finite isospin chemical potential $\mu_I$}
We will be using chiral perturbation theory to describe the pion-condensed phase
of QCD at finite $\mu_I$ and zero temperature. $\chi$PT is a low-energy effective theory
for QCD based on the (global) symmetries and degrees of freedom. It provides
a correct model-independent description as long as one is inside its
domain of validity~\cite{weinberg,gasser1,gasser2}. The effective Lagrangian that describes the low-energy degrees of freedom of QCD (pions, kaons, eta) can be written in a low-energy expansion,
\bqa
{\cal L}&=&{\cal L}_2+{\cal L}_4+...\;,
\eqa
where the subscript $n$ denoted the order in the expansion. The expansion parameter
can be written as ${M\over4\pi f}$ with $M$ being a relevant mass or momentum scale
and $f$ is the pion-decay constant. The leading-order Lagrangian is
the nonlinear sigma model, which for two flavors reads
    \bqa
{\cal L}_{2}&=&{1\over4}f^2\langle\nabla^{\mu} \Sigma^{\dagger} \nabla_{\mu}\Sigma\rangle
+{1\over4}f^2\langle\chi^{\dagger}\Sigma+\Sigma^{\dagger}\chi\rangle\; ,
\eqa
where $\langle\rangle$ means trace in flavor space and 
\bqa
\Sigma=e^{i{\phi_a\tau_a\over f}}\;,
\hspace{2cm}
\chi=2B_0{\rm diag}(m_u,m_d)\;,
\eqa
with $\phi_a$ being the Goldstone bosons fields, $m_{u,d}$ are the quark masses, and
$B_0$ is the related to the quark condensate in the vacuum via $\langle\bar{\psi}\psi\rangle=-f^2B_0$.
The covariant derivative is
\bqa
\label{cov}
\nabla_{\mu} \Sigma\equiv
\partial_{\mu}\Sigma-i\left [v_{\mu},\Sigma \right]\;,
\eqa
with 
$v_{\mu}=\delta_{\mu0}(\mbox{$1\over3$}\mu_B\mathbb{I}+\mbox{$1\over2$}\mu_I\tau_3)$.
Note that the results will be independent of $\mu_B$ since the unit matrix 
$\mathbb{I}$ commutes with everything, this reflects that the
mesons have zero baryon number.

The most general ansatz for the (normalized) ground state can after some simplifcations
be written as~\cite{son}
\bqa
    \Sigma_{\alpha}&=&
    \mathbb{I}\cos\alpha+i\tau_2\sin\alpha
=e^{i\alpha\tau_2}
    \;,
\eqa
which simply is a rotation in flavor space of the vacuum state $\mathbb{I}$ 
by an angle $\alpha$. The leading-order thermodynamic potential is
  \bqa
\Omega_0&=&-f^2B_0(m_u+m_d)\cos\alpha-{1\over2}f^2\mu_I^2\sin^2\alpha\;.
\label{oo}
  \eqa
We see that there is a competition between the two terms in Eq.~(\ref{oo}), the first term prefers the vacuum state $\alpha=0$, while the second term prefers $\alpha={1\over2}{\pi}$.
The optimum is found by balancing these two terms in the 
thermodynamic potential,
\bqa
\cos\alpha&=&{m_{\pi,0}^2\over\mu_I^2}\;,
\hspace{1cm}\mu_I^2\geq m_{\pi,0}^2\;, 
\\ 
\alpha&=&0\;,\hspace{1cm}\mu_I^2<m_{\pi,0}^2
\;,
\eqa
where $m_{\pi,0}^2=B_0(m_u+m_d)$ is the tree-level pion mass.
Thus there is phase transition from the vacuum to a pion-condensed phase
at a critical $\mu_I$, $\mu_I^c=\pm m_{\pi,0}$. To determine the order of the transition,
one can construct a Ginzburg-Landau energy functional by expanding
the thermodynamic potential $\Omega_0$ around $\alpha=0$,
\bqa
\Omega_0&=&-f^2m_{\pi,0}^2+{1\over2}f^2[m_{\pi,0}^2-\mu_I^2]\alpha^2-{1\over24}f^2
\left[m_{\pi,0}^2-4\mu_I^2
\right]\alpha^4+{\cal O}(\alpha^6)\;.
\eqa
We define a critical chemical isospin potential $\mu_I^c$ 
when the order-$\alpha^2$ term
vanishes, i.e. $\mu_I^c=\pm m_{\pi,0}$. Since the coefficient of
the order-$\alpha^4$ term is positive for $\mu_I=\mu_I^c$, the transition is second order. Note that all thermodynamic quantities are independent of $\mu_I$
for all $\mu_I^2<m_{\pi,0}^2$ implying that e.g. the isospin density vanishes in the same region
and not only for $\mu_I=0$. This is an example of the Silver-Blaze property~\cite{cohen0}.


\section{Phase diagram for three flavors}
We next consider the phase diagram for three-flavor QCD at finite $\mu_I$ and $\mu_S$
including electromagnetic effects.
If we couple $\chi$PT to dynamical photons, the Lagrangian contains a few
extra terms at leading order in the low-energy expansion~\cite{gasser89},
\bqa
\nonumber
{\cal L}_2&=&-{1\over4}F_{\mu\nu}F^{\mu\nu}+
{1\over4}f^2\langle\nabla_{\mu}\Sigma\nabla^{\mu}\Sigma^{\dagger}
\rangle
+{1\over4}f^2
\langle\chi^{\dagger}\Sigma+\Sigma^{\dagger}
\chi\rangle
+C\langle Q\Sigma Q\Sigma^{\dagger}\rangle
\\ &&
+{\cal L}_{\rm gf}+{\cal L}_{\rm ghost}
\;,
\label{lagel}
\eqa
with $\chi=2B_0{\rm diag}(m_u,m_d,m_s)$ and $\Sigma=e^{i{\phi_a\lambda_a\over f}}$.
The new term $C\langle Q\Sigma Q\Sigma^{\dagger}\rangle$ is responsible for the 
tree-level mass splitting between the charged and neutral pions, and it also contributes 
to the tree-level mass splitting between the charged and neutral kaons.
The inclusion of electromagnetic effects also implies that the phases with charged
meson condensates are superconductors and that the massless degree of freedom
(the Goldstone bosons) is eaten up by the now massive photon via the Higgs mechanism.
The term $v_{\mu}$ in the covariant derivative Eq.~(\ref{cov}) is replaced by
\bqa
v_0&=&\mbox{$1\over3$}(\mu_B-\mu_S)\mathbb{I}+\mbox{$1\over2$}\mu_{K^{\pm}}
\lambda_Q+{1\over2}\mu_{K^0}\lambda_K\;,
\hspace{1cm}
v_i=0
\;,
\eqa
where
\bqa
\mu_{K^{\pm}}&=&{1\over2}\mu_I+\mu_S\;,
\hspace{2cm}
\mu_{K^0}=-{1\over2}\mu_I+\mu_S\;,\\
\lambda_Q&=&\lambda_3+{1\over\sqrt{3}}\lambda_8\;,
\hspace{1.8cm}
\lambda_K=-\lambda_3+{1\over\sqrt{3}}\lambda_8\;.
\eqa
In analogy with the two-flavor case, we expect onset of charged kaon condensation
when $\mu_{K^{\pm}}^2=m_{K^{\pm}}^2$ and neutral kaon condensation when
$\mu_{K^{0}}^2=m_{K^{0}}^2$.
The corresponding ans\"atze for the ground states are~\footnote{One could imagine
multiple condensates in parts of the $\mu_I$--$\mu_S$ plane, but that is ruled
out by actual calculations. The angles $\beta$ and $\gamma$ are the rotation angles
of the quark condensate into a charged or neutral kaon condensate, respectively.}
\bqa
\Sigma_{\beta}=e^{i\beta\lambda_5}\;,
\hspace{3cm}
\Sigma_{\gamma}=e^{i\gamma\lambda_7}\;.
\eqa
The thermodynamic potential in the different phases can then be computed
as functions of the chemical potentials. 
For example, in the charged kaon condensed phase, the thermodynamic potential is 
  \bqa
  \nonumber
\Omega_0&=&
-f^2B_0(m_u+m_s)\cos\alpha
-{1\over2}f^2
\left[\mu_{K^{\pm}}^2-\Delta m^2_{\rm EM} \right]\sin^2\alpha
\;,
\eqa
where $\Delta m^2_{\rm EM}={2Ce^2\over f^2}$ is the splitting between the 
charged and neutral kaons due to electromagnetism. It follows that
the transition takes place exactly at $\mu_{K^{\pm}}^2=m_{K^{\pm}}^2=B_0(m_u+m_d)+{2Ce^2\over f^2}$ as expected.

\begin{figure}[htb!]
         \centering
         \includegraphics[width=\columnwidth]{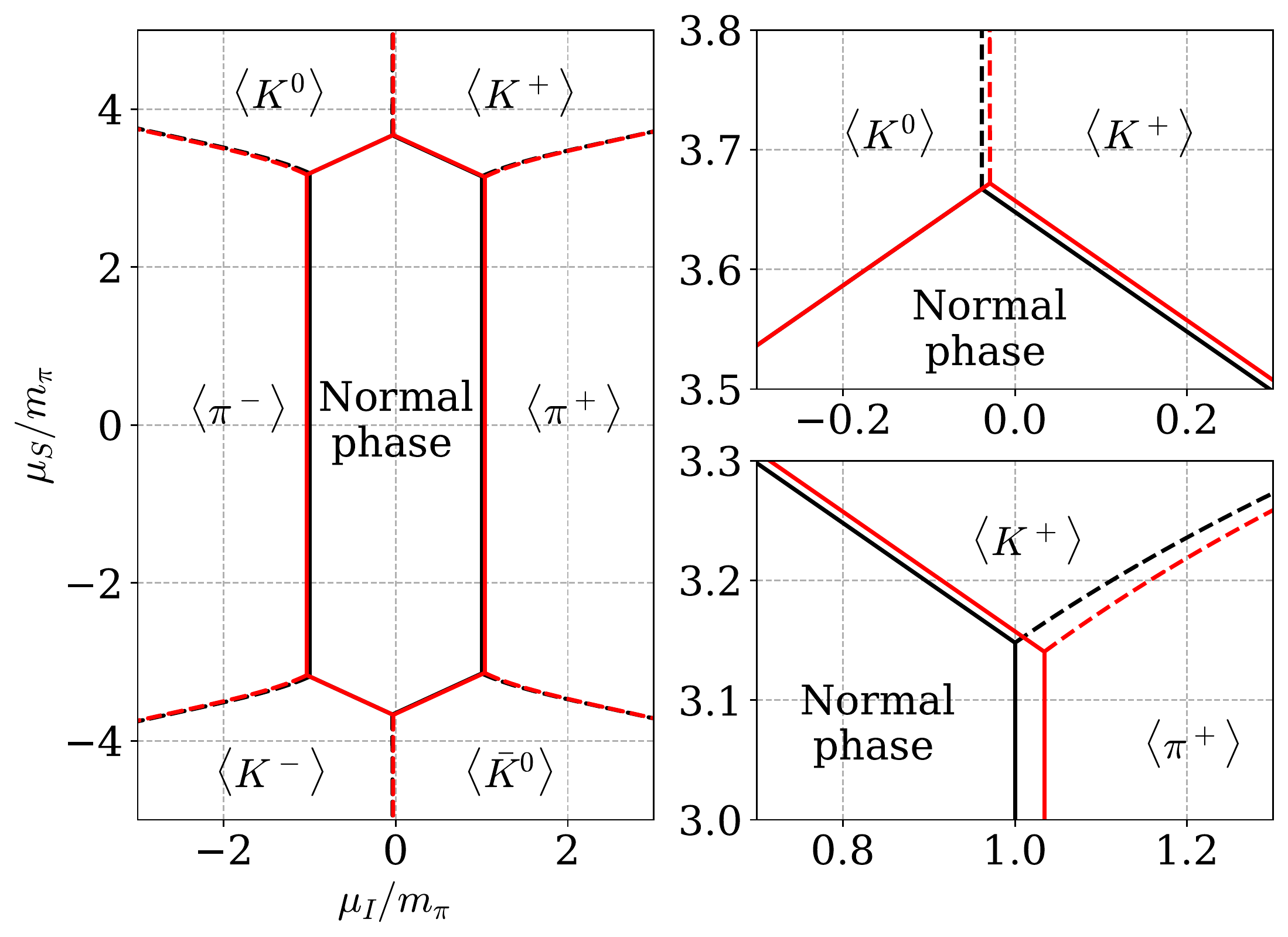}
        \caption{Left panel shows the phase diagram as predicted by $\chi$PT 
        in the $\mu_I$--$\mu_S$-plane. 
In the right panel, we have zoomed in on the triple points. See main text for details.
Fig. from Ref.~\cite{johnsrud}.}
        \label{diagram}
\end{figure}

For each value of $(\mu_I,\mu_S)$, we find the
phase with the lowest value of $\Omega_0$ (or largest pressure). 
This phase wins and we can map out the phase diagram 
in this manner. The result is shown in the left panel of Fig.~\ref{diagram}.
The black lines are the transition lines without electromagnetic
interactions and the red lines are with electromagnetic interactions. 
The former was first obtained by Kogut and Toublan~\cite{kogut33} in the isospin limit.
The phases with charged meson condensation become Higgs phases upon including
electromagnetic effects. The tree-level mass of the photon is $m_A=ef\sin\alpha$.
The transitions from the
normal phase to a meson-condensed phase is always second order
with mean-field exponents in the $O(2)$ universality class. The transitions
between the various condensed phases are always first order
and involve the competition between the order parameters
of the different phases. As we cross the transition lines, the order parameters as well as the isospin and strangeness densities, $n_I$ and $n_S$
jump discontinuously. 
The small offset of the dashed vertical lines is due to 
the mass difference between the charged and neutral 
kaons, which is both due to $\Delta m_\text{EM}\neq0$, and $m_u\neq m_d$.
These contributions, however, pull in opposite directions, 
as we see in the phase diagram. The contribution due to the difference in
quark masses adds to the mass of the $K^0/\bar{K}^0$ meson, which is why the black transition
line between the kaon condensate is to the left of the $\mu_I = 0$ line, 
while the electromagnetic contribution adds to the mass of the charged kaon,
which is why the red line is between these two lines. 
The partition function in the normal phase is independent of the two chemical 
potentials $\mu_I$ and $\mu_S$, which again is the Silver Blaze
property~\cite{cohen0}. 
In the right panels, we have zoomed in on the triple points.
Upper panel shows the intersection of the normal, neutral kaon condensed and charged kaon
condensed phases, while the lower shows the intersection of the normal, pion condensed, and charged kaon condensed phases.

\section{${\cal O}(p^4)$ calculation of thermodynamic potential}
I will next sketch the NLO calculation of the thermodynamic potential. For simplicity
I consider two flavors in the isospin limit, $m_u=m_d=m$.
The thermodynamic potential can be calculated in a low-energy expansion,
  \bqa\nonumber
\Omega&=&\Omega_0+\Omega_1+...,
\eqa
where $\Omega_n$ is the order-${\cal O}(p^{2n+2})$ contribution.
The term $\Omega_1$ receives contributions from the one-loop graphs of
${\cal L}_2^{\rm quadratic}$ and counterterms coming from 
${\cal L}_4^{\rm static}$.
The relevant terms are
\bqa
\nonumber
\mathcal{L}_2^{\text{quadratic}}
&=&\mbox{$1\over2$}
(\partial_{\mu}\phi_{a})
(\partial^{\mu}\phi_{a})+\mu_{I}\cos\alpha(\phi_{1}\partial_{0}\phi_{2}-\phi_{2}
\partial_{0}\phi_{1})\\ \nonumber
&&-\mbox{$1\over2$}\left [(m^{2}_{\pi,0}\cos\alpha-\mu_{I}^{2}\cos2\alpha)\phi_{1}^{2}
\right.
+(m^{2}_{\pi,0}\cos\alpha-\mu_{I}^{2}\cos^{2}\alpha)\phi_{2}^{2}\\
&&\left.+(m^{2}_{\pi,0}\cos\alpha+\mu_{I}^{2}\sin^{2}\alpha)\phi_{3}^{2}\right]
\label{l2}
\;,\\ \nonumber
  \mathcal{L}_{4}^{\text{static}}&=&(l_{1}+l_{2})\mu_{I}^{4}\sin^4\alpha
  +l_{4}m^{2}_{\pi,0}\mu_{I}^{2}\cos\alpha\sin^{2}\alpha
+(l_{3}+l_{4})m^{4}_{\pi,0}\cos^{2}\alpha
+h_1m_{\pi,0}^4\;,
\\ &&
\;,
\label{l4}
\eqa
where $l_1$--$l_4$ and $h_1$ are bare couplings.
They are related to the renormalized couplings ${l_i^r}$ and $h_i^r$
via
$l_i=l_i^r(\Lambda)+{\gamma_i\Lambda^{-2\epsilon}\over2(4\pi)^2}\left[{1\over\epsilon}+1\right]$ and
$h_i=h_i^r(\Lambda)+{\delta_i\Lambda^{-2\epsilon}\over2(4\pi)^2}\left[{1\over\epsilon}+1\right]$, where
$\gamma_i$, $\delta_i$ are coefficients and $\Lambda$ is the renormalization scale
in the $\overline{\rm MS}$ scheme. Since $\delta_1=0$, $h_1=h_1^r$ and does not run.
Performing the Gaussian integral over the quantum fields $\phi_a$ 
in dimensionsal regularization using 
Eq.~(\ref{l2}), we obtain a divergent contribution to $\Omega_1$.
The divergences are cancelled by adding the quartic terms from Eq.~(\ref{l4})
and renormalizing the couplings by making replacing the bare couplings with the
renormalized ones.
The final result is
\bqa\nonumber
\Omega_0+\Omega_1&=&-f^{2}m^{2}_{\pi,0}\cos\alpha-\frac{1}{2}f^{2}\mu_{I}^{2}
\sin^{2}\alpha
\\ && \nonumber
-\frac{1}{4(4\pi)^{2}}\left [\frac{3}{2}-\bar{l}_{3}
  +4\bar{l}_{4}
  +\log\left({m_{\pi,0}^2\over \tilde{m}_2^2}\right)
+  2\log\left({m_{\pi,0}^2\over m_3^2}\right)
\right ]
m_{\pi,0}^{4}\cos^{2}\alpha
\\&& \nonumber
-\frac{1}{(4\pi)^{2}}
\left [{1\over2}+\bar{l}_{4}
  +  \log\left({m_{\pi,0}^2\over m_3^2}\right)
\right ]
m^{2}_{\pi,0}\mu_{I}^{2}\cos\alpha\sin^{2}\alpha
\\ &&
\nonumber
-\frac{1}{4(4\pi)^{2}}\left [1
  +\frac{2}{3}\bar{l}_{1}+\frac{4}{3}\bar{l}_{2}
  +  2\log\left({m^2_{\pi,0}\over m_3^2}\right)
\right ]\mu_{I}^{4}
\sin^{4}\alpha
\\ &&
-{1\over(4\pi)^2}\bar{h}_1m_{\pi,0}^4
+V_{{1},\pi^+}^{\text{fin}}+V_{{1},\pi^-}^{\text{fin}}
\;,
\label{effpotnlo}
\eqa
where $\tilde{m}_2^2=m_{\pi,0}^2\cos\alpha$, 
$m_3^2=m_{\pi,0}^2+\mu_I^2\sin^2\alpha$, and 
$V_{{1},\pi^+}^{\text{fin}}+V_{{1},\pi^-}^{\text{fin}}$ are two complicated
finite terms that must be evaluated numerically.
Finally, $\bar{l}_i$ and $\bar{h}_i$ are, up to a prefactor, equal to $l_i^r$
and $h_i^r$ at the scale $\Lambda=m_{\pi,0}$.
Using Eq.~(\ref{effpotnlo}) one can show that the phase transition takes place
at $\mu_I^c=m_{\pi}$, where the physical pion mass now includes 
radiative corrections~\cite{gasser1}, see Eq.~(\ref{mpi}) below.
The parameters $\bar{l}_i$ are determined by experiment and $\bar{h}_1$
estimated by model calculations.
and the parameters $m_{\pi,0}^2=2B_0m$ and $f$ can be found by inverting the one-loop
relations using the experimental values for the pion mass and the pion decay constant,
\bqa
\label{mpi}
m_{\pi}^2=m^2_{\pi,0}\left[1-{m^2_{\pi,0}\over2(4\pi)^2f^2}\bar{l}_3\right]\;,
\hspace{1cm}
f_{\pi}^2
=f^2\left[1+{2m^2_{\pi,0}\over(4\pi)^2f^2}\bar{l}_4\right]\;.
\eqa

\section{Condensates}
\begin{figure}[htb!]
    \centering
        \includegraphics[scale=0.5]{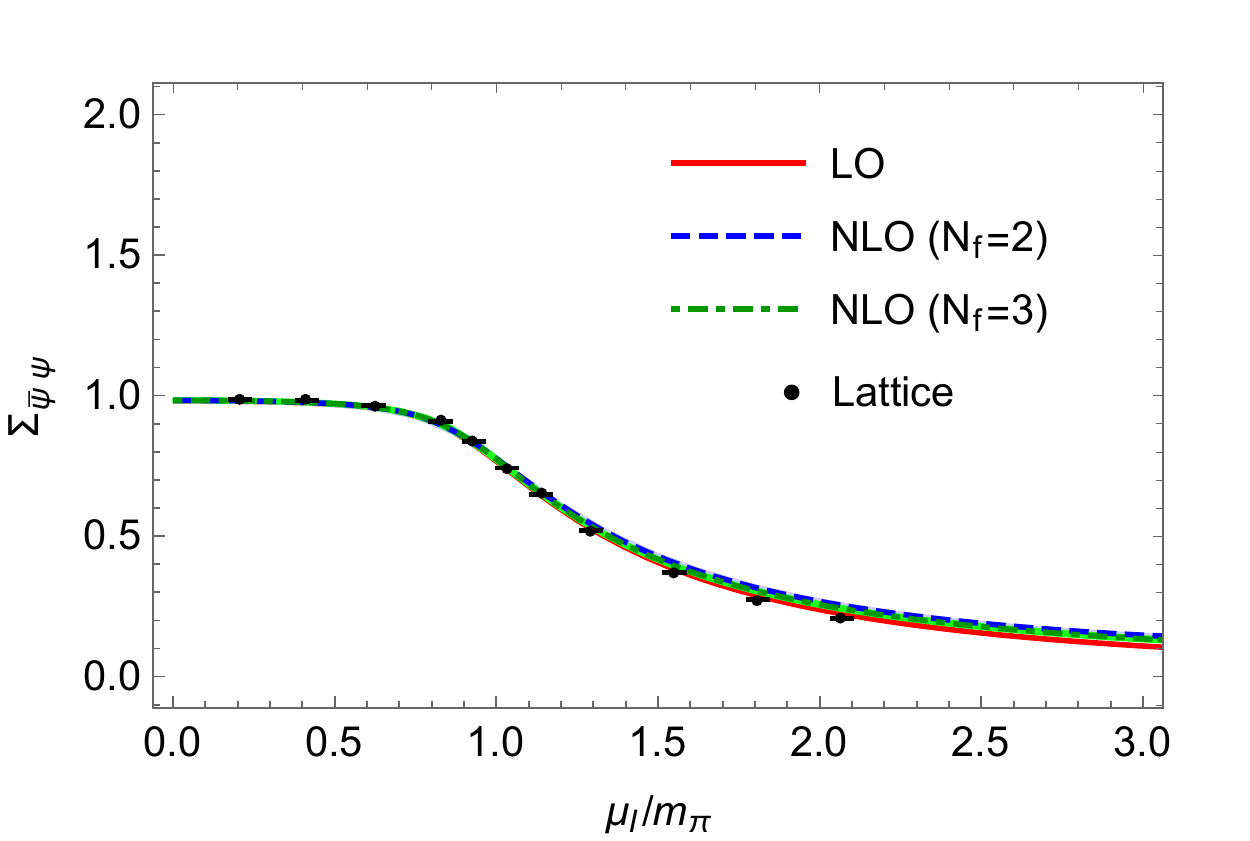}
  \includegraphics[scale=0.5]{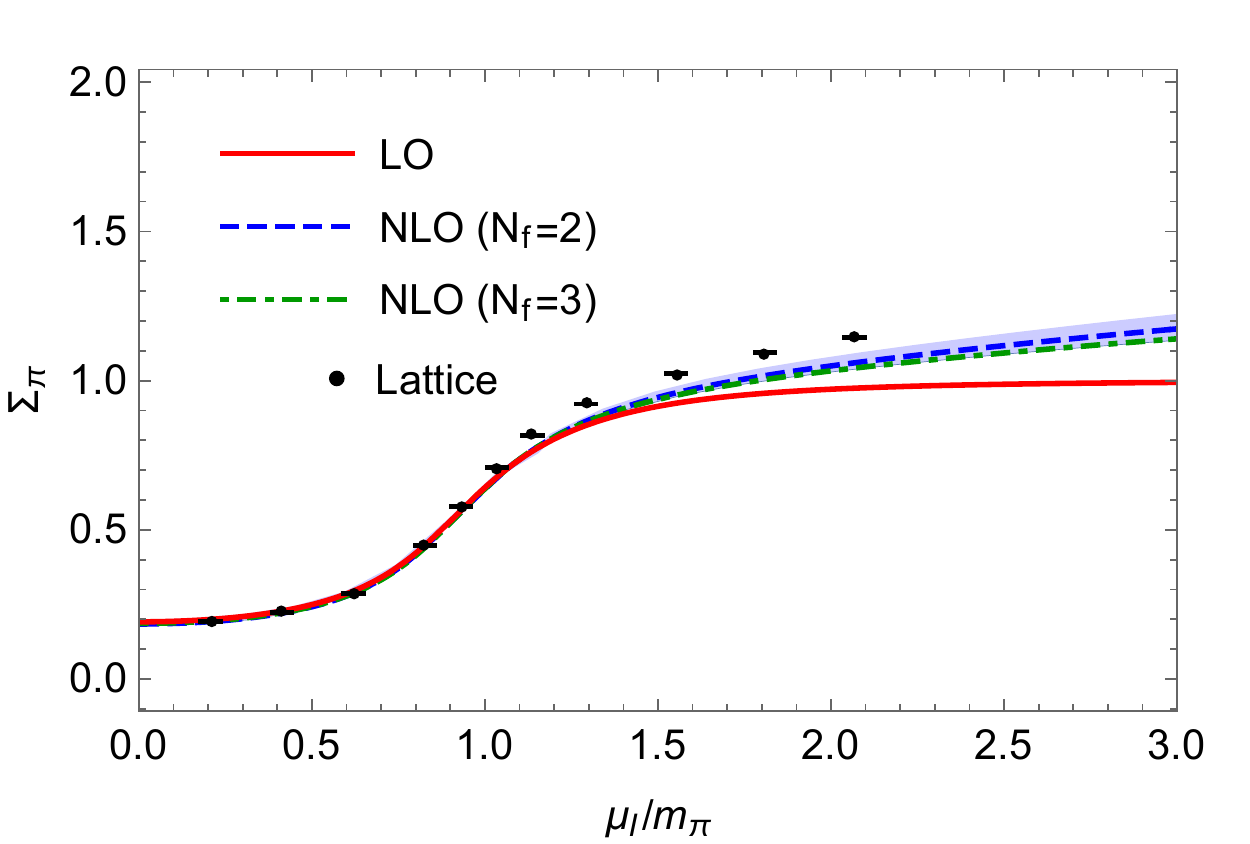}
  \caption{$\Sigma_{\bar{\psi}\psi}$ (left panel) and $\Sigma_{\pi}$ (right panel) as functions of
  $\mu_I/m_{\pi}$ at zero temperature and finite source $j=0.00517054m_{\pi}$. Fig. taken from Ref.~\cite{martin}.}
\label{condie}
\end{figure}

In order to obtain the light quark and pion condensates, we need to calculate the
thermodynamic potential $\Omega$ with sources $m$ and $j$, where the latter is a
pionic source. 
The former has already been 
included in the calculations I have shown and it is also straightforward to
include a pionic source $j$ in the calculations. For example, in the two-flavor
expression for the thermodynamic potential, one simply makes the replacements
$m\cos\alpha\rightarrow m\cos\alpha+j\sin\alpha$
and 
$\bar{h}_1m_{\pi,0}^4=\bar{h}_1(2B_0m)^2\rightarrow\bar{h}_1[(2B_0m)^2+(2B_0j)^2]$~\cite{martin}.
Once these replacements are made, the condensates are given by
\bqa
\label{c1}
\langle \bar{\psi}\psi\rangle_{\mu_I}
={1\over2}{\partial \Omega\over \partial m}
=-f^2B_0\cos\alpha+...\;,
\hspace{1cm}
\langle \pi^+\rangle_{\mu_I}={1\over2}{\partial \Omega\over \partial j}
=-f^2B_0\sin\alpha+...\;,
\label{c2}
\eqa
where I on the the right-hand side have written explicitly the tree-level contributions.
The subscript $\mu_I$ on the expectation values indicate that they depend on the
chemical potential. Instead of plotting the condensates directly, we define
the normalized deviations as 
\bqa
\label{deviation}
\Sigma_{\bar{\psi}\psi}=-\frac{2m}{m_{\pi}^{2}f_{\pi}^{2}}\left[\langle\bar{\psi}\psi\rangle_{\mu_{I}}^{}-\langle\bar{\psi}\psi\rangle_{0}^{j=0}\right]+1\;,
\hspace{1cm}
\label{deviation2}
\Sigma_{\pi}=-\frac{2m}{m_{\pi}^{2}f_{\pi}^{2}}\langle\pi^{+}\rangle^{}_{\mu_{I}}\;.
\eqa
At tree level, Eq.~(\ref{c2}) shows the rotation of the 
quark condensate into a pion condensate.
Equivalently, from Eq.~(\ref{deviation2}), the deviations at tree level satisfy
$\Sigma_{\bar{\psi}\psi,\rm tree}+\Sigma^2_{\pi,\rm tree}=1$.
This interpretation no longer holds beyond ${\cal O}(p^2)$.
In the left panel of Fig.~\ref{condie}, we show $\Sigma_{\bar{\psi}\psi}$
as a function of ${\mu_I\over m_{\pi}}$ at leading order (red line)~\footnote{The leading order result is the same for two and three flavors.} 
and next-to-leading order
for two flavors (blue line) and three flavor (green line).
The data points are from the lattice simulations of Ref.~\cite{gergo,gergo1,gergo2,gergo3}.
In the right panel, we show $\Sigma_{\pi}$ in the same approximations.
We note that the difference between  $\Sigma_{\bar{\psi}\psi}$ in the various approximations is very 
small and they all agree very well with the lattice data points. Regarding 
$\Sigma_{\pi}$, we notice that it is nonzero
for $\mu_I<m_{\pi}$, which simply reflects that the curves shown are for nonzero 
pion source, $j=0.00517054m_{\pi}$ 
The $U(1)_{I_3}$-symmetry is therefore broken explicitly
for all values of $\mu_I$. Comparing the various approximations and lattice data,
it is evident that including the ${\cal O}(p^4)$ corrections results in a substantially better agreement between $\chi$PT and simulations.
All the numerical results have been obtained by using the same physical meson masses
as well as $f_{\pi}$
as in the lattice simulations. This requires that we invert the relations  
Eq.~(\ref{mpi}) to obtain the values for the bare mass $m$ and the bare pion decay constant $f$ using the experimental values of $\bar{l}_i$.

\section{Acknowledgements}
I would like to thank Prabal Adhikari, Martin Mojahed, and Martin Kj{\o}llesdal
Johnsrud for collaboration. I would also like to thank the organizers of
the XVth Quark confinement and the Hadron spectrum conference for a very interesting meeting.

\bibliography{references}

\end{document}